\documentclass[journal]{IEEEtran}
\usepackage{times}  
\usepackage{helvet}  
\usepackage{courier}  
\usepackage[hyphens]{url}  
\usepackage{graphicx} 
\usepackage[numbers]{natbib} 
\usepackage{caption} 
\usepackage{algorithm}
\usepackage{algpseudocode}
\usepackage{newfloat}
\usepackage{listings}
\usepackage{subcaption}

\usepackage{xcolor}
\usepackage{colortbl}
\usepackage{multirow}
\usepackage{placeins}

\ifCLASSINFOpdf

\else

\fi

\usepackage{ifthen}

\usepackage{amsfonts,amssymb,amsmath,amsthm}
\usepackage{xspace}
\usepackage{xcolor}
\usepackage{tikz}
\usepackage{subcaption}

\newcommand{\roni}[1]{\textbf{Roni: #1}}
\newcommand{\tuple}[1]{\left\langle{#1}\right\rangle}
\newcommand{\prefix}{\textit{w}\xspace}
\newcommand{\budget}{B\xspace}
\newcommand{\lnstwo}{MAPF-LNS2\xspace}
\newcommand{\lnsone}{MAPF-LNS\xspace}
\newcommand{\confprop}{ConflictProportion\xspace}
\newcommand{\revconfprop}{ReverseConflictProportion\xspace}
\newcommand{\conflicts}{\textit{conf}\xspace}
\newcommand{\revconflicts}{\textit{revconf}\xspace}

\newcommand{\PID}{PID\xspace}
\newcommand{\mab}{Multi-Arm-Bandit\xspace}

\newcommand{\plan}[2]{%
    \ifthenelse{\equal{#1}{DONE}}%
        {}%
        {\textcolor{red}{TODO: #2}}%
}

\usepackage{bibentry}
\begin{document}

\title{Budget Allocation Policies for Real-Time Multi-Agent Path Finding}

\author{Raz Beck\textsuperscript{\rm 1},
    Roni Stern\textsuperscript{\rm 1},
    Amir Shapiro\textsuperscript{\rm 1}}

\maketitle

\begin{abstract}
Multi-Agent Path finding (MAPF) is the problem of finding paths for a set of agents such that each agent reaches its desired destination while avoiding collisions with the other agents. This problem arises in many robotics applications, such as automated warehouses and swarms of drones.
Many MAPF solvers are designed to run offline, that is, first generate paths for all agents and then execute them.
In real-world scenarios, waiting for a complete solution before allowing any robot to move is often impractical. Real-time MAPF (RT-MAPF) captures this setting by assuming that agents must begin execution after a fixed planning period, referred to as the planning budget, and execute a fixed number of actions, referred to as the execution window.
This results in an iterative process in which a short plan is executed, while the next execution window is planned concurrently.
Existing solutions to RT-MAPF iteratively call windowed versions of MAPF algorithms in every planning period, without explicitly considering the size of the planning budget. 
We address this gap and explore different policies for allocating the planning budget in windowed versions of MAPF-LNS2, a state-of-the-art MAPF algorithm. 
Our exploration shows that the baseline approach in which all agents draw from a shared planning budget pool is ineffective in challenging scenarios. 
Instead, policies that intelligently distribute the planning budget among agents are able to solve more problem instances in less time. 
\roni{Let's go over the last bit of the abstract again at the end}
Since no single policy consistently outperforms the others across all parameter settings, we propose an oracle-like approach that selects the best-performing policy for each parameter configuration. For example, on the Random map, the baseline achieves the highest success rate in 44 percent of the experimental configurations, whereas the oracle-like approach attains the highest success rate across all configurations.
\end{abstract}


\IEEEpeerreviewmaketitle

\section{Introduction}

\IEEEPARstart{M}{ulti} -Agent Path finding (MAPF) problem involves finding collision-free paths for multiple agents navigating a shared environment from given initial positions to their designated targets. MAPF has many applications including automated warehouse, traffic management, robotic swarm control and more. 
From the computational complexity point of view, the problem is NP-Hard to solve optimally for various optimization criteria~\cite{yu2013structure,surynek2021multi} and even NP-Hard to solve at all in directed graphs~\cite{nebel2020computational}. 
Nevertheless, many fast MAPF algorithms exists and can scale to solve MAPF problems with thousands of robots very quickly. 

\plan{DONE}{What is real-time MAPF problem, why it is important}
In this work, we focus on solving MAPF under \emph{real-time constraints}, which means that the agents must commit to perform their next sequence of 
actions, termed the \emph{prefix}, within a fixed, strict time budget, termed the \emph{planning budget}. We refer to this problem as \emph{Real-Time MAPF (RT-MAPF)} ~\cite{Devon2018rthsmapf}. 
 The real-time constraints in RT-MAPF are common in realistic applications, especially where path planning is but one part of a larger system. For example, in an automated warehouses path planning must occur in tandem with target allocation, low-level robotic control, and other considerations. The robots in a warehouse may not wait in their place until planning is done, and planning must be interleaved with execution.

\plan{DONE}{Existing solutions: solve a windowed version of MAPF}
Interleaving planning and execution has been studied in the context of MAPF.
One approach is to use a fast rule-based or learned policy~\cite{okumura2022priority,skrynnik2024learn}. 
While fast, this approach tends to produce lower quality solutions due to their myopic nature. 
A more common alternative, called ``windowed planning'', involves iteratively planning for a limited horizon, ignoring collisions that may occur after the horizon~\cite{li2021lifelong}. 
Windowed planning, however, is not guaranteed to return a solution within the available planning budget. 
Morag et al.~\cite{morag2023adapting} proposed a framework for handling cases where the planner is not able to return a solution in time. To handle such cases,  they proposed several \emph{fail policies} that take a partial solution and modify it such that conflicts are avoided. Zhang et al.~\cite{zhang2024planning} proposed a similar framework, where planning is done during execution, as well as a more sophisticated fail policy. 
In both studies, the primary emphasis was placed on the general real time MAPF framework and fail policies. The planning itself was agnostic to the available planning budget beyond halting when it is exhausted. 
\textbf{The key challenge we consider in this work is to design a RT-MAPF algorithm that explicitly considers the available planning budget}, especially when we do not have enough budget to find a feasible solution.

\plan{DONE}{Our contribution: budget allocation policies in PrP and MAPF LNS2}
\textbf{Our first contribution} is a framework for budget-aware RT-MAPF algorithms. Our framework is built on the \lnstwo algorithm~\cite{Li2022lns2}, which is a state-of-the-art MAPF algorithm. The \lnstwo algorithm operates in two distinct stages. The first stage calculates the optimal path for each agent independently, ignoring other agents' paths, which may result in conflicting paths. The second stage iteratively repairs these paths until a feasible, conflict-free solution is achieved. 
In our RT-MAPF framework, we run a modified version of \lnstwo
which uses a \emph{Budget Allocation Policy (BAP)}  to decide how much planning budget to allocate to a selected neighborhood of agents. When the budget allocated to a neighborhood is exhausted, we stop planning for that neighborhood and keep the agents' current plans unchanged. If a solution was found and more planning budget exists, we invest it to improve the incumbent solution using \lnsone (an anytime algorithm similar to \lnstwo designed to improve existing valid solutions). 

\textbf{Our second contribution} is several possible BAPs. These BAPs include one that allocates more budget to neighborhoods of agents that conflict more; one that allocates less budget to such  neighborhoods;one that is inspired by PID controllers~\cite{nise2019control}; and one based on Multi-Armed Bandit (MAB) method~\cite{katehakis1987multi}. 

\textbf{Our third contribution} is an extensive experimental evaluation of our framework and these BAPs over a set of standard MAPF maps and a range of planing budgets. Our results show that no single BAP provides the best solution in all cases. We analyze empirically where each works best, and show that by choosing the best one for each configuration, we obtain significant advantage in terms of the ability to solve problems within a reasonable time. For example, for one of the problem configurations we observe that the best policy solves 100\% of the configuration scenarios while the best baseline solves only 76\%.

\section{Background and Problem Definition}

The classical MAPF problem~\cite{stern2019multi} involves a graph $G = (V, E)$ and a set $\mathcal{K}$ of $k$ agents representing the robots. Each agent $i \in \{1,\ldots,k\}$ has a designated source vertex $s_i$ and target vertex $g_i$. We define a path $\pi = (v_1, \ldots, v_{|\pi|})$ as a sequence of vertices where consecutive vertices are either identical (representing a \emph{wait} action) or connected by an edge (representing a \emph{move} action).
Time is assumed to be discrete, the duration of every move is a single time step, and the cost of a path $\pi$, denoted $C(\pi)$ is the number of time steps needed to traverse the path ($|\pi| - 1$). 
 
A MAPF \emph{solution} $\Pi$ is an assignment mapping each robot $i$ to a conflict-free path $\Pi(i)$ from $s_i$ to $g_i$. We focus on two conflict types: \emph{vertex conflicts}, where robots move to the same vertex simultaneously, and \emph{swapping conflicts}, where robots exchange positions in a single move. 
Sum of costs (SOC) and makespan are two common MAPF solution cost functions where $\textit{SOC}(\Pi) = \sum_{\pi \in \Pi} C(\pi)$ and $\textit{Makespan}(\Pi)=\max_{\pi \in \Pi} C(\pi)$. Minimizing \textit{SOC} reduces the total movement of all robots, while minimizing \textit{Makespan} focuses on ensuring the last robot reaches its goal as quickly as possible.  
\plan{DONE}{(1-2 paragraphs at most) Describe MAPF Algorithms: PrP, PIBT, LNS2}

\subsection{MAPF Algorithms}
Finding optimal solutions to MAPF is computationally intractable~\cite{nebel2020computational} and thus less suitable for solving MAPF in real-time for a large number of agents. 
Therefore, in this work, we focus on two well-known, suboptimal and fast MAPF algorithms, namely,  Priority Inherence Backtracking (PIBT) \cite{okumura2022priority} and \lnstwo~\cite{Li2022lns2}.

PIBT~\cite{okumura2022priority} is an extremely fast MAPF algorithm that finds a solution by iteratively choosing the next step for all agents until the agents reach their targets. 
For every time step, agents are assigned random priorities and select their optimal next moves based on this hierarchy.
If an agent lacks a valid, conflict-free move, it employs a Priority Inheritance technique combined with backtracking. 
Under certain conditions, PIBT guarantees each agent eventually visits its target, albeit not necessarily at the same time. LaCAM~\cite{okumura2023lacam} and LaCAM*~\cite{okumura2024engineering} are complete MAPF algorithms that invoke a systematic search mechanism on top of PIBT, allowing it to ``backtrack'' if it reaches a dead-end instead of failing.

\lnstwo is a state-of-the-art MAPF algorithm based on Large Neighborhood Search~\cite{Li2022lns2}.
It has an \emph{initial planning} phase and a \emph{neighborhood search} phase. 
The initial planning phase finds a fast initial solution, which may contain conflicts. 
 
The neighborhood search phase of \lnstwo is designed to resolve the remaining conflicts and improve the overall solution quality. 
It works by iteratively choosing a fixed number of agents, referred to as a \emph{neighborhood}, 
and finding new paths for these agents that avoid conflicts between the agents in the neighborhood. 
\lnstwo uses multiple heuristic methods for choosing the agents in the neighborhood, including conflict-based methods and random selection. 
To find new paths for the agents in a neighborhood, \lnstwo can use any MAPF algorithm. 
The common implementation of \lnstwo is to use Prioritized Planning (PrP)~\cite{pp-old}. 
In PrP, agents are ordered by some priority and plan sequentially, with each agent computing a path to its target that avoids the paths already planned for the higher-priority agents. 
\plan{DONE}{How SIPPS works} One way to compute a single-agent path that avoids other agents' plans is to use A* on a time-expanded graph, where a node is a vertex-time step pair.
SIPP~\cite{phillips2011sipp} improves the efficiency of this single-agent path finding by grouping collision-free time steps into safe intervals and planning over these intervals instead of individual time states. 
Standard implementation of \lnstwo uses SIPPS~\cite{Li2022lns2}, which extends SIPP by also trying to minimize the conflicts with  paths outside the currently planned neighborhood.  

In online versions of the MAPF problem either agents appear and disappear over time~\cite{vsvancara2019online} or agents receive new targets over time~\cite{li2021lifelong}. 
The latter type of online MAPF, referred to as \emph{Lifelong MAPF (LMAPF)} received significant attention in the literature due to its practical applications in multi-agent pick-up and delivery (MAPD)~\cite{Ma2017}.  
Rolling Horizon Collision Resolution (RHCR) is a common framework for solving LMAPF problems. 
RHCR accepts two parameters, the \emph{execution window} $w$ and the 
planning \emph{horizon} $h$, and alternates between planning and execution. 
During planning, a MAPF algorithm is used to find a path for each agent such that the agents do not conflict in the first $h$ steps. 
The agents then commit and execute the first $w$ steps in the found paths. Conflicts are guaranteed to be avoided during execution as long as $h\geq w$.  
Limiting the planning horizon is intended to speed up the search, as well as account for the uncertainty that stems from not knowing the future targets of each agent. 
\plan{DONE}{SOTA for LMAPF}
To find paths during planning, RHCR requires a ``windowed'' MAPF algorithm, i.e., one that ignores conflicts after the planning horizon. 

\subsection{Real-Time Single Agent Search}
Real-time single-agent search is class of search problems in which a single-agent must commit to an action in constant time~\cite{KORF1990rths}. 
This constant time is referred to as the \emph{planning budget} and is usually measured by the number of nodes expanded. 
Reaching the goal may therefore require interleaving planning and execution. 
A common approach to solve real-time single agent search problems is to use Real-Time Heuristic Search (RTHS) algorithms~\cite{Sharon2014exponential,bulitko2006learning, hernandez2012avoiding, koenig2009comparing}.  
RTHS algorithms rely on a heuristic function to guide the search while interleaving planning and execution. 
These algorithms perform iteratively the following steps: (1) run a local lookahead search from the current state $s$ until the planning budget is exhausted; (2) update its heuristic function based on the searched states; (3) select a sequence of actions to perform from $s$; and (4) execute these actions and update the current state accordingly. 
This process repeats until the goal is roached or some pre-defined stopping criteria is met. 
RTHS algorithm differ in how they perform the lookahead search, how they update the heuristic function, and how they select the sequence of actions to perform.

Recent approaches adapt RTHS to MAPF in different ways. Sigurdson et al.~\cite{sigurdson2018multi} propose Bounded Multi-Agent A* (BMAA*), where each agent runs an RTHS algorithm separately, while treating other agents as dynamic obstacles. 
Veerapaneni et al.~\cite{veerapaneni2025windowed} proposed the WinC-MAPF framework, which runs an RTHS algorithm on the joint configuration space of all agents. 
Their framework executes the standard RTHS cycle on this joint space but maintains computational tractability within the planning bound $B$ by limiting the planning horizon and use a windowed version of MAPF algorithm to suggest the next (joint) action to perform. They also identified disjoint agent groups and update the heuristic accordingly. 

\section{Problem Definition}

The Real-Time MAPF (RT-MAPF) problem we consider in this work is defined by a tuple $\tuple{G,\mathcal{K},s,g, \budget, \prefix}$ where $\tuple{G,\mathcal{K},s,g}$ is a classical MAPF problem, 
\budget is the planning budget allowed for every planning period,  
and \prefix is the number of steps each agent must execute before the next planning period. 
A RT-MAPF algorithm is called in the beginning of every planning period, and must output a \emph{solution prefix}, which is a mapping $\Pi_\prefix$ that maps every agent to a path of size $\prefix$. 
The agents must be able to execute every solution prefix, so an RT-MAPF algorithm must ensure every solution prefix it returns does not have a vertex or swapping conflict.

In general, \budget may quantify any form of computational resources that is limited, e.g., time or memory. 
In this work, we follow the RTHS literature and have \budget be the maximal number of single-agent search nodes expanded in every planning period.  
Relating search nodes to runtime is reasonable in our context, since most MAPF algorithms we consider rely on performing multiple single-agent path finding searches. 
RT-MAPF is different from LMAPF. 
In LMAPF, there is no explicit planning budget and when an agent reaches its target it may receive a new one. Also, in LMAPF there is no need for all agents to reach their targets at the same time. 

\subsection{Existing Solutions}
\plan{DONE}{(1-2 paragraphs at most) existing approaches and motivation}

One may consider using RHCR as-is to solve RT-MAPF by setting the planning horizon to be sufficiently small so that paths are found within the planning budget, or incrementally increasing the planning horizon until the planning budget is exhausted~\cite{li2021anytime}. 
This solution may lead to conflicts, since if the planning horizon is set to be smaller than the prefix, conflicts may occur during execution. On the other hand, setting the planning horizon to be equal to or greater than the prefix may be too large, exhausting the planning budget before finding a conflict-free solution within the planning horizon.

Morag et al.~\cite{morag2023adapting} identified this limitation of RHCR and proposed a framework for LMAPF that can handle the real-time constraints we consider. 
They referred to cases where a valid plan could not be found by RHCR in time as a \emph{planning failure}~\cite{morag2023adapting,zhang2024planning} and explored several policies to address it. 
Specifically, they proposed to extract a \emph{partial solution} from the planner that has failed and apply a \emph{fail policy} to synthesize conflict-free paths from it. 
A partial solution $\Pi$ is a mapping of agents to paths such that every $\Pi(i)$ starts in agent $i$'s current location but does not necessarily end up in agent $i$'s target. 
A \emph{fail policy} is a fast algorithm that accepts a \emph{partial solution} and outputs a MAPF solution in which the agents do not conflict within the execution window. 
A trivial fail policy is to have all agents stay in their place. A more effective yet simple fail policy, called \emph{IStay}, is to have conflicting agents stay in their place while letting other non-conflicting agents continue to move according to the returned partial solution. They also proposed a more sophisticated fail policy called IAvoid, but its benefits were relatively minor. 

The Planning and Improving while Executing (PIE) framework~\cite{zhang2024planning} 
builds on Morag et al.'s framework, emphasizing that planning should occur during execution as opposed to halting the system for planning after every execution window. 
In addition, they proposed several improvements including 
a more sophisticated fail policy. Using PIE yielded impressive performance in both LMAPF and MAPF scenarios. 
 
\section{Finding Useful Partial Solutions}
Both frameworks -- Morag et al.'s and PIE -- require a planner that is able to return partial solutions in case of planning failures. 
Morag et al.~\cite{morag2023adapting} adapted PrP~\cite{pp-old} for this purpose. In PIE~\cite{zhang2024planning}, they first run a fast suboptimal MAPF algorithm, namely LaCAM*, to find an initial solution and then run \lnsone to improve it. 
If a planning failure occurs during the initial planning period, they select the best node explored so far by LaCAM*.
Otherwise, if a planning failure occurs, then the partial solution returned is the initial, possibly low-quality solution found by LaCAM*. 
All these previously proposed methods do not directly consider the planning budget beyond using it as a cutoff. 
Next, we propose alternative budget-aware methods to adapt \lnstwo to return useful partial solutions.

\section{Budget Allocation Policies for \lnstwo}

\begin{figure}[h]
    \centering
    \begin{tikzpicture}[scale=0.8]
        \foreach \x in {1,...,7} {
            \foreach \y in {0,...,3} {
                \draw (\x-1, \y) rectangle (\x, \y+1);
            }
        }
        \filldraw[black] (3, 1) rectangle (4, 3);

        \filldraw[blue] (3.5, 0.5) circle (0.4);
        \node[white] at (3.5, 0.5) {$s_{8}$};
        \filldraw[blue] (3.5, 3.5) circle (0.4);
        \node[white] at (3.5, 3.5) {$s_9$};

        \draw[blue] (4.5, 2.5) circle (0.4);
        \node[black] at (4.5, 2.5) {$t_{8}$};

        \draw[blue] (4.5,1.5) circle (0.4);
        \node[black] at (4.5, 1.5) {$t_9$};

        \foreach \y [count=\i from 0] in {0,...,3} {
            \filldraw[gray] (0.5, \y+0.5) circle (0.4);
            \node[white] at (0.5, \y+0.5) {$s_{\i}$};
        }
        \foreach \y [count=\i from 4] in {0,...,3} {
            \filldraw[gray] (1.5, \y+0.5) circle (0.4);
            \node[white] at (1.5, \y+0.5) {$s_{\i}$};
        }

        \foreach \y [count=\i from 4] in {0,...,3} {
             \draw[gray] (6.5, \y+0.5) circle (0.4);
             \node[black] at (6.5, \y+0.5) {$t_{\i}$};
        }
        \foreach \y [count=\i from 0] in {0,...,3} {
             \draw[gray] (5.5, \y+0.5) circle (0.4);
             \node[black] at (5.5, \y+0.5) {$t_{\i}$};
        }

    \end{tikzpicture}
    \caption{A difficult MAPF configuration that will be hard to solve without budget allocation}
    \label{fig:bottleneck_config}
\end{figure}

\plan{DONE}{Baseline approach to get partial solutions with LNS}

\lnstwo relies heavily on its neighborhood selection policies, which are not perfect. 
Thus, committing all the planning budget to a selected neighborhood may be detrimental in our real-time setting. 
For example, consider the MAPF problem illustrated in Fig.~\ref{fig:bottleneck_config}. 
Committing all the planning budget to plan for agents 0-7 will be ineffective since 
the initial paths of agents 8 and 9 conflict and without replanning they will block all paths of the other agents. 
We consider limiting the planning budget allocated to each neighborhood to avoid such over-commitment issues. 

We explored various forms of a \emph{neighborhood budget policy} for choosing how much planning budget to allocate for every neighborhood. Given a specific neighborhood, these policies determine the planning budget allocated to it.
Inspired by PIE ~\cite{zhang2024planning}, we ran PIBT simultaneously to \lnstwo, choosing the initial solution that minimizes the farthest away agent from its goal. Comparing the amount of planning budget spent by PIBT and \lnstwo is problematic, as PIBT does not search in the same search space. Thus, we made the simplifying assumption that the computational cost of running PIBT is zero. This can be justifies by either running PIBT in parallel to \lnstwo on a different processor, or empirically, since PIBT is usually extremely fast. After obtaining an initial solution we then ran \lnsone to try and improve the partial solution with the remaining planning budget. Following the execution of the partial solution, if any agent has not yet reached its goal, the process iterates, generating a new partial solution for the subsequent execution window. This process is shown in the Algorithm.~\ref{alg:budget_allocation_framework}. In the Algorithm.~\ref{alg:lns2_budget_policy}, the function iteratively refines a the partial solution $\pi$ to resolve agent conflicts within a strict global budget. In each loop, the algorithm selects a neighborhood of agents and queries the Budget Allocation Policy (BAP) to determine a the planning budget cap ($B_\mathcal{N}$) for that subgroup, which is immediately deducted from the global remaining budget ($B_{rem}$). Agents within the neighborhood are sequentially replanned via SIPPS, where the local budget is consumed at a unit cost per search node expansion. After the replanning phase, any unused portion of the allocated local budget is reclaimed and added back to the global pool, and the process repeats until the solution is conflict-free or the total budget is exhausted.

\begin{algorithm}[t]
\caption{Budget Allocation Framework}
\label{alg:budget_allocation_framework}
\begin{algorithmic}[1]
\Require $\tuple{G, \mathcal{K}, s, g, \budget, \prefix, BAP}$
\State $\Pi \gets s$
\State $t \gets 0$
\While{$\Pi[-1] \neq g$}
    \State \textbf{in parallel:}
    \State \quad $\pi_1 \gets \textsc{PIBT\_Initial}(G, \Pi, \prefix)$
    \State \quad $\pi_2 \gets \textsc{LNS2\_Initial}(G, \Pi, \prefix)$
    \State $\pi_2, \budget_{\text{rem}} \gets 
        \Call{LNS2\_Improve}{\pi_2, \budget,  \prefix, BAP}$
    \State $\pi \gets \Call{MinMaxDistance}{\pi_1, \pi_2}$
    \If{$\budget_{\text{rem}} > 0$}
        \State $\pi \gets \Call{LNS1}{\pi, \budget_{\text{rem}},\prefix}$
    \EndIf
    \State $\Pi \gets \Pi \oplus \pi$
    \State $t \gets t + 1$
\EndWhile
\State \Return $\Pi$
\end{algorithmic}
\end{algorithm}

\begin{algorithm}
    \caption{LNS2 search improvement process with budget allocation policy}
    \label{alg:lns2_budget_policy}
    \begin{algorithmic}[1]
        \Function{LNS2\_Improve}{$\pi, \budget, \prefix, BAP$}:
            \State $B_{rem} \gets \budget$
            
            \While{$\Call{Conflicts}{\pi} \land B_{rem}$}
                \State $\mathcal{N} \gets \Call{Neighborhood\_Selection}{\pi}$
                
                \State $B_\mathcal{N} \gets \Call{BAP}
                {\mathcal{N}, \pi, B_\mathcal{N}, \prefix}$
                \State $B_{rem} \gets B_{rem}-B_\mathcal{N}$
                \For{$agent \in \mathcal{N}$}
                    \State $\pi, B_\mathcal{N} \gets \Call{SIPPS}{agent, \pi, B_\mathcal{N}}$
                \EndFor
                
                \State $B_{rem} \gets B_{rem}+B_\mathcal{N}$
            \EndWhile
            
            \State \Return $\pi, B_{rem}$
        \EndFunction
    \end{algorithmic}
\end{algorithm}

\paragraph{Neighborhood budget policies}

The baseline neighborhood budget policy corresponds to allocating all the available planning budget to the current neighborhood, until it either finds paths for its constituent agents or fails. We call this the \emph{Shared} neighborhood budget policy. 

As an alternative, we propose the following neighborhood budget allocation policies
\paragraph{Conflict Proportional Budget}
The first policy we introduce is the \emph{Conflict Proportional Budget}, referred to as \confprop. \confprop assigns an amount of budget proportional to the amount of conflicts the agents in the chosen neighborhood are involved in. 
Formally, let $\conflicts(i)$ be the number of conflicts agent $i$ is involved in the incumbent solution, 
$\mathcal{N}$ be the set of agents in the chosen neighborhood, $\mathcal{K}$ be the set of all agents, 
and $B$ be the total planning budget still available. 
\confprop allocates to neighborhood $\mathcal{N}$ the following amount of budget, denoted by $B(\mathcal{N})$ as shown in Eq.~\ref{eq:CPB}:
\begin{equation}
B(\mathcal{N})=B \cdot \sum_{i\in \mathcal{N}} \conflicts(i)/\sum_{j\in \mathcal{K}} \conflicts(j) 
\label{eq:CPB}
\end{equation}
A limitation of using $B(\mathcal{N})$ to allocate budget for neighborhoods is that in hard instances with many conflicts the budget allocated to each neighborhood may be too small to find any plan. For example, if the execution window $w$, then a neighborhood $\mathcal{N}$ will need at least $w\cdot |\mathcal{N}|$ nodes to find paths even if no conflicts occur. To mitigate this, we imposed the following lower bound on the budget that can be given to a neighborhood $\mathcal{N}$, denoted $B_L(\mathcal{N})$ and computed as shown in Eq.~\ref{eq:CPB_lowerbound}:
\begin{equation}
B_L(\mathcal{N}) = \frac{|\mathcal{N}|(|\mathcal{N}|+1)}{2}\cdot w 
\label{eq:CPB_lowerbound}
\end{equation}
We explored other methods for computing this lower bound and this performed best. In conclusion, for a neighborhood $\mathcal{N}$ the \confprop allocates a budget of $\max(B(\mathcal{N}),B_L(\mathcal{N}))$ for planning.
\paragraph{Reversed Conflict Proportional Budget}
In addition, we propose \revconfprop, a budget allocation policy designed to assign more budget to agents with fewer conflicts. Due to the way SIPPS operates, an agent terminates its search once a feasible path is found; any unused budget is then returned to the shared budget pool for future use. This policy prioritizes finding feasible paths for less constrained agents first, with the goal of simplifying the planning problem for more challenging agents.
\revconfprop allocates budget following Eq.~\ref{eq:RCPB}:
\begin{equation}
B(\mathcal{N}) = B \cdot \frac{\sum_{i\in \mathcal{N}} \revconflicts(i)}{\lvert \mathcal{N} \rvert  \cdot \sum_{j\in \mathcal{K}} \conflicts(j) }
\label{eq:RCPB}
\end{equation}
Where $\revconflicts(i)$ is set by Eq.~\ref{eq:RCPB_single_agent}:
\begin{equation}
\revconflicts(i) =\sum_{j\in \mathcal{K}} \conflicts(j) - \conflicts(i) 
\label{eq:RCPB_single_agent}
\end{equation}
This method ensures that neighborhoods with fewer conflicts receive a larger budget, and vice versa.
\paragraph{PID}
Another alternative policy is inspired by the \PID controller. While the \PID controller is designed to minimize error in a feedback control system by considering the error itself, its derivative, and its integral, we adapt these principles into our problem. Much like the \confprop policy, the \PID allocates higher budget to agents who have more conflicts in their plan. The main difference between the two policies is that the \PID does not output a proportional number to the remaining overall budget, but instead assigns an absolute value.
\begin{flushleft}
\begin{equation}
B(\mathcal{N})=\prefix \cdot \sum_{i\in \mathcal{N}}
    \frac{k_{p}}{\conflicts(i) + 1}
    + k_{d} \cdot \Delta \conflicts(i)
    + k_{i} \cdot C_{i}
\label{eq:PID_conflict}
\end{equation}
\end{flushleft}
Equation~\ref{eq:PID_conflict} defines the conflict-driven \PID, where $k_{p}$, $k_{d}$, and $k_{i}$ are the \PID parameters. The term $\Delta\text{conflicts}(i)$ denotes the change in the number of conflicts in agent $i$’s current path relative to its path at the beginning of the planning period. Finally, $C_{i}$ is a non-negative integer associated with each agent that is incremented by one each time the agent is selected as part of a neighborhood.
In classical control theory, \PID parameters are typically synthesized from the system's governing equations to minimize error. However, due to the absence of an these equations in our problem, we determined the parameters via empirical optimization. Following an evaluation of the parameter space, the chosen values were identified as $k_{p}=1, k_{d}=1.5, k_{i}=1$.
\paragraph{Multi-Arm-Bandit}
The final method we propose is an online learning mechanism that selects which BAP to apply to a chosen neighborhood. To achieve this, we incorporate a Multi-Armed Bandit (MAB) framework that includes the previously described budget allocation policies: \confprop, \revconflicts, and \PID. The MAB maintains a weight vector, where each weight corresponds to one of the candidate policies. When a neighborhood is selected for replanning, the probability of choosing a particular policy is determined by its current weight. After the neighborhood search concludes, whether successfully completed or prematurely terminated, the weight of the selected policy is updated. This update captures the effectiveness of the search process. If the search is successful, meaning that the newly computed paths contain fewer conflicts than the previous ones, the weight of the selected policy is increased. Otherwise, if no improvement is achieved, the weight is decreased.
The weight of the chosen policy is updated according to Eq.~\ref{eq:MAB}:
\begin{equation}
Policy_{w_updated} = const \cdot is_improved + (1 - const) \cdot Policy_w
\label{eq:MAB}
\end{equation}
Here, $Policy_w$ denotes the current weight of the selected policy. The parameter $const$ is a constant that determines the influence of the latest outcome on the updated weight. The variable $is_improved$ is a binary indicator that equals 1 if the newly generated paths contain fewer conflicts than the previous paths, and 0 otherwise.

The example given in Fig.~\ref{fig:bottleneck_config} highlights the advantage of using the \revconfprop budget policy. Because agents 0–7 exhibit the most conflicts, neighborhoods containing them are allocated only a small budget fraction. This preserves resources for agents 8 and 9, allowing the solver to find conflict-free paths and resolve the bottleneck once they are selected.

For distributing budget inside the neighborhood, we utilizes a Shared policy baseline: the entire allocated budget is initially made available to the first agent, and any residual budget is sequentially passed to the subsequent agent.
We also investigated alternative strategies, including a fixed budget policy for neighborhoods and inside them for distributing the planning budget. However, as these methods yielded negligible performance differences compared to the proposed approaches, they have been omitted to maintain the clarity and focus of the analysis. 

\section{Experimental Results}

We included in our experiments all the algorithms and budget policies described above.
This includes \lnstwo~\cite{li2021anytime} with the baseline \emph{shared}, \PID, \confprop, \revconfprop and \mab policies; 
denoted Shared, PID, CPB, RCPB and MAB. 
As a baseline, we also only ran PIBT to get a fast initial solution, which satisfies our real-time constraints. 

\begin{figure*}[t]
    \centering  
    \begin{subfigure}[b]{0.15\textwidth}
        \centering
        \includegraphics[width=\textwidth]{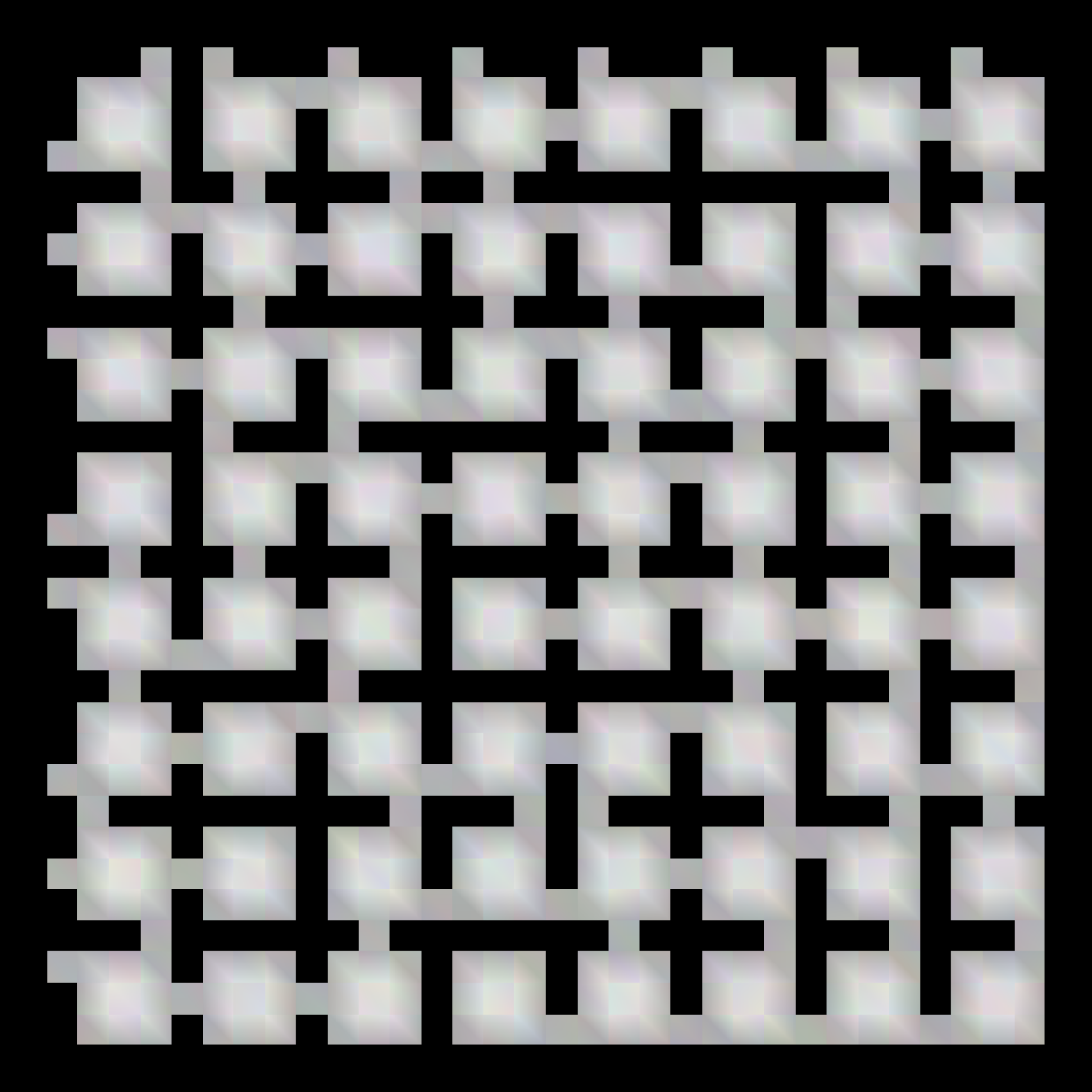}
        \caption{Room Map}
        \label{fig:room-map}
    \end{subfigure}   
    \begin{subfigure}[b]{0.15\textwidth}
        \centering
        \includegraphics[width=\textwidth]{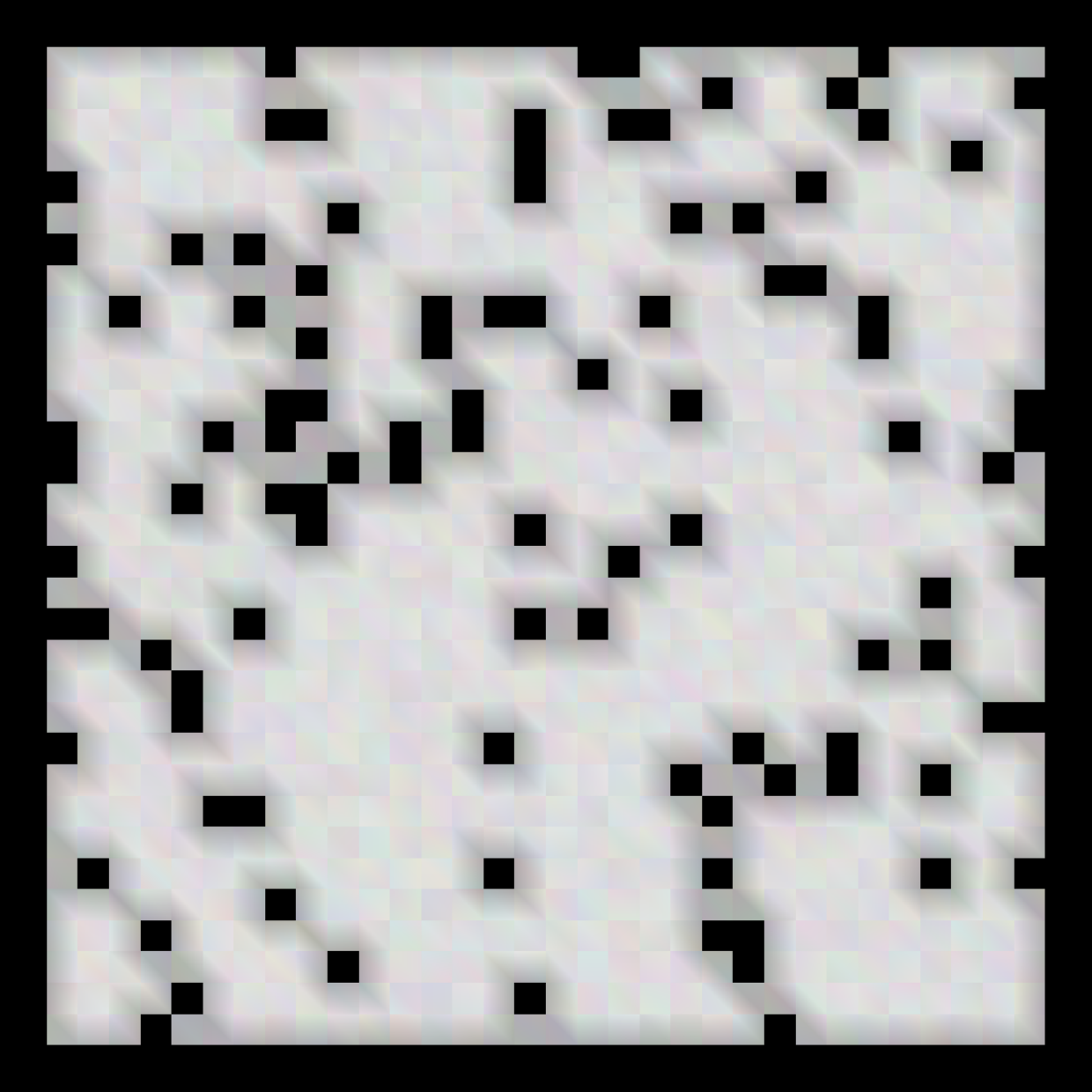}
        \caption{Random Map}
        \label{fig:random-map}
    \end{subfigure}
    \begin{subfigure}[b]{0.15\textwidth}
        \centering
        \includegraphics[width=\textwidth]{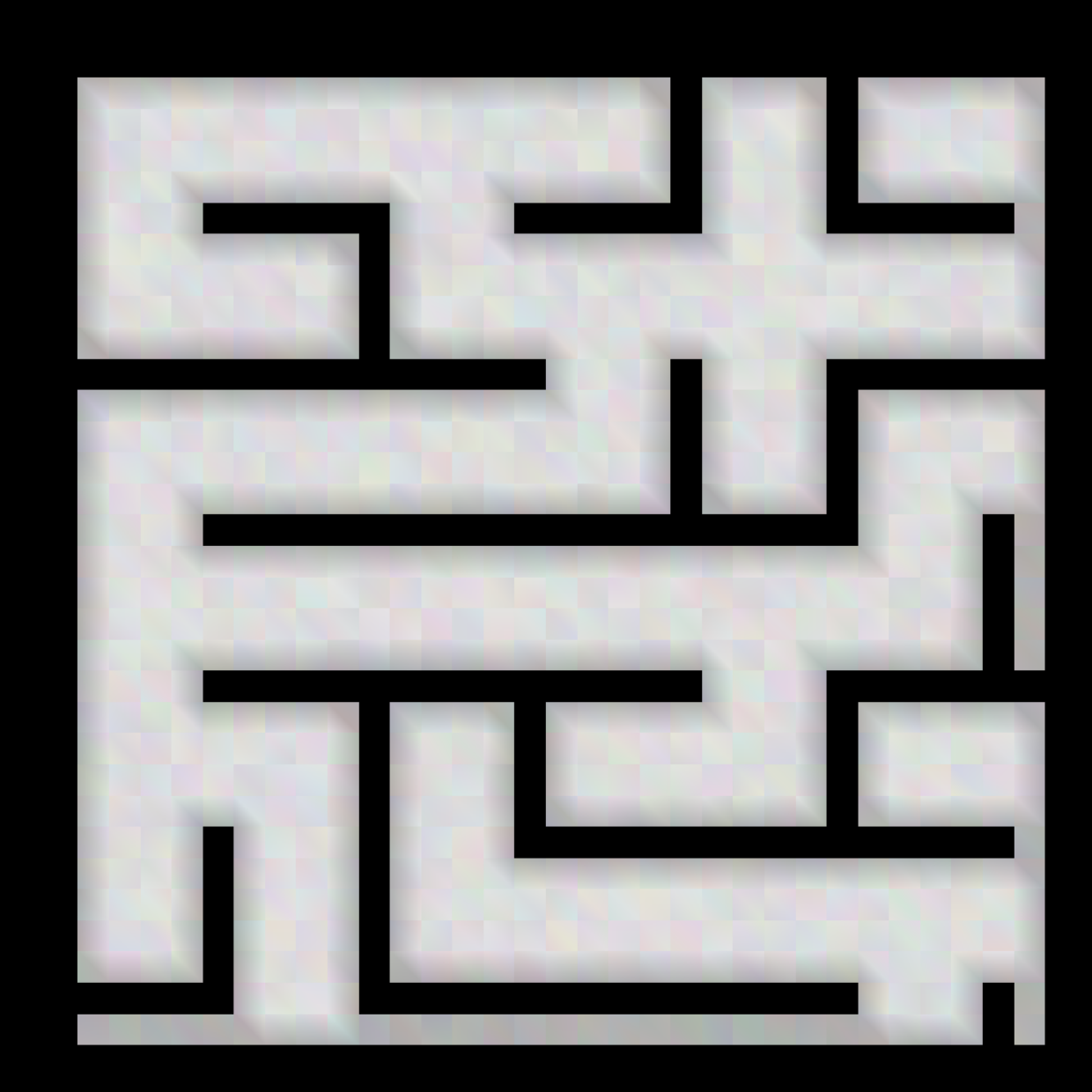}
        \caption{Maze Map}
        \label{fig:maze-map}
    \end{subfigure} \quad
    \caption{Experiment's maps}
    \label{fig:maps}
\end{figure*}

This enables PIBT to output an initial solution without spending any budget, leaving the complete budget available for \lnsone.
For the \mab policy we set the const to $0.01$ and the starting weights of each policy to $1.00$.
We evaluated the proposed budget policies and baselines experimentally on different types of grids from the standard MAPF benchmark~\cite{stern2019multi}. 
Specifically, we ran experiments on the following grids \texttt{room 32-32-4}, \texttt{random 32-32-10} and \texttt{maze 32-32-4}, denoted 
here as Room, Random and Maze, shown in Fig.~\ref{fig:maps}. 

[[Raz: Morag Framework maybe isn't necesary anymore because we built out own.]]
To meet the real-time MAPF requirements, all algorithms ran within Morag et al.'s~\cite{morag2023adapting} robust MAPF framework. As explained earlier, this framework builds on RHCR~\cite{li2021lifelong}, using a limited planning horizon and committing to perform the first $w$ steps in the resulting solution. This process repeats until all agents are at their targets. 
Since the algorithms we consider do not ensure completeness in reasonable time, we imposed a maximum number of steps after which the experiment is declared failed if the agents are not at their targets. We set this upper bound on the solution makespan to be 100. 
The experiments ran on different combinations of execution windows and budget per agent, while the number of agents in the problem was fixed for each map. The execution windows were [5,10,15,20,25,30,35], the budgets per agent were [5,10,15,20,25,50,80,100,150,200,250,300,350,400] and the number of agents for the maps Room, Random and Maze were [120,300,27] respectively. 
The planning budget is based on the idea that each agent is equipped with a computing unit that contributes computational power per time step.

The results indicate that no single policy is universally superior. Therefore, we propose an oracle algorithm called "Single Policy Oracle", denoted as SPO, which selects the policy with the highest success rate for each parameter composition. Consequently, our analysis primarily focuses on comparing SPO against the baselines. Additionally, we include a theoretical upper bound, denoted simply as Oracle, constructed by selecting the best-performing policy for every individual scenario within each parameter composition.

\subsection{Results 1: Success Rate}
We first assess performance using the success rate. A scenario is considered a success if a solution is found within the makespan's upper bound.
The percentage of parameter pairs where each policy achieved the highest success rate is shown in Table~\ref{tab:policies-success-rate-percentages}. In this table we observe that no policy consistently yielded the best results. Although PIBT had the lowest win rate across all maps, CPB performed best in Room (52\%) and Maze (71\%), while MAB attained the highest percentage (54\%) in the Random map. Table~\ref{tab:policies-success-rate-ratios} shows the ratio of the success rate of each individual policy against the success rate of the SPO and the Oracle. The parameter pairs in this table represent a spectrum of difficulty, ranging from strict budget constraints relative to the execution window, to scenarios with abundant resources allocated for larger windows. In this table, we observe that no single policy has a great advantage over the other policies, indicating that there is no universally superior policy.

\begin{table}[]
\centering
\begin{tabular}{|c|c|c|c|c|c|c|}
\hline
       & PIBT & Shared & CPB & RCPB & PID & MAB \\ \hline
Room   & 11    & 49     & 52  & 47   & 41  & 42  \\ \hline
Random & 12     &44    &51 &44 &46 &51 \\ \hline
Maze   & 21   & 67     & 69  & 68   & 64  & 61  \\ \hline
\end{tabular}
\caption{Percentage of the 98 parameter configurations where each policy yielded the best performance.}
\label{tab:policies-success-rate-percentages}
\end{table}

\begin{table*}[]
\centering
\begin{tabular}{|c|c|c|c|c|c|c|c|c|c|c|}
\hline
Map & Prefix & Budget & PIBT & Shared & CPB & RCPB & PID & MAB & SPO & Oracle \\ \hline
\multirow{9}{*}{Room} & \multirow{3}{*}{5} & 5 & \textbf{1} & 0.51 & 0.51 & 0.51 & 0.51 & 0.62 & \textbf{1} & 1.02 \\ \cline{3-11} 
 &  & 50 & 0.80 & \textbf{1} & \textbf{1} & \textbf{1} & 0.84 & 0.80 & \textbf{1} & \textbf{1} \\ \cline{3-11} 
 &  & 200 & 0.83 & 0.96 & 0.96 & 0.96 & \textbf{1} & \textbf{1} & \textbf{1} & 1.04 \\ \cline{2-11} 
 & \multirow{3}{*}{20} & 5 & \textbf{1} & \textbf{1} & \textbf{1} & \textbf{1} & 0.82 & 0.91 & \textbf{1} & 1.55 \\ \cline{3-11} 
 &  & 50 & 0.77 & 0.96 & \textbf{1} & 0.91 & 0.91 & 0.87 & \textbf{1} & 1.09 \\ \cline{3-11} 
 &  & 200 & 0.74 & 0.96 & 0.96 & 0.96 & \textbf{1} & \textbf{1} & \textbf{1} & 1.04 \\ \cline{2-11} 
 & \multirow{3}{*}{35} & 5 & \textbf{1} & 0.83 & 0.83 & 0.83 & 0.83 & 0.83 & \textbf{1} & 1.33 \\ \cline{3-11} 
 &  & 50 & 0.83 & 0.83 & 0.94 & \textbf{1} & 0.67 & 0.67 & \textbf{1} & 1.28 \\ \cline{3-11} 
 &  & 200 & 0.61 & 0.91 & 0.87 & 0.91 & 0.83 & \textbf{1} & \textbf{1} & 1.09 \\ \hline
\multirow{9}{*}{Random} & \multirow{3}{*}{5} & 5 & \textbf{1} & 0.91 & 0.91 & 0.91 & 0.96 & 96 & \textbf{1} & 1.04 \\ \cline{3-11} 
 &  & 50 & \textbf{1} & \textbf{1} & \textbf{1} & \textbf{1} & \textbf{1} & \textbf{1} & \textbf{1} & 1.10 \\ \cline{3-11} 
 &  & 200 & 0.91 & 0.96 & 0.96 & 0.96 & 0.96 & \textbf{1} & \textbf{1} & 1.09 \\ \cline{2-11} 
 & \multirow{3}{*}{20} & 5 & 0.75 & 0.92 & 0.92 & 0.96 & \textbf{1} & 0.92 & \textbf{1} & \textbf{1} \\ \cline{3-11} 
 &  & 50 & 0.88 & \textbf{1} & 0.96 & \textbf{1} & 0.92 & 0.96 & \textbf{1} & \textbf{1} \\ \cline{3-11} 
 &  & 200 & 0.96 & \textbf{1} & \textbf{1} & \textbf{1} & \textbf{1} & \textbf{1} & \textbf{1} & \textbf{1} \\ \cline{2-11} 
 & \multirow{3}{*}{35} & 5 & 0.50 & 0.90 & \textbf{1} & 0.90 & 0.90 & 0.90 & \textbf{1} & 1.05 \\ \cline{3-11} 
 &  & 50 & 0.78 & 0.87 & 0.91 & 0.91 & \textbf{1} & 0.96 & \textbf{1} & 1.09 \\ \cline{3-11} 
 &  & 200 & 0.87 & 0.96 & \textbf{1} & 0.96 & \textbf{1} & \textbf{1} & \textbf{1} & 1.09 \\ \hline
\multirow{9}{*}{Maze} & \multirow{3}{*}{5} & 5 & 0.8 & \textbf{1} & \textbf{1} & \textbf{1} & \textbf{1} & \textbf{1} & \textbf{1} & 1.20 \\ \cline{3-11} 
 &  & 50 & 0.76 & \textbf{1} & \textbf{1} & \textbf{1} & \textbf{1} & \textbf{1} & \textbf{1} & 1.12 \\ \cline{3-11} 
 &  & 200 & 0.71 & \textbf{1} & \textbf{1} & \textbf{1} & \textbf{1} & \textbf{1} & \textbf{1} & \textbf{1} \\ \cline{2-11} 
 & \multirow{3}{*}{20} & 5 & \textbf{1} & 0.82 & 0.82 & 0.82 & 0.82 & 0.82 & \textbf{1} & 1.27 \\ \cline{3-11} 
 &  & 50 & 0.75 & \textbf{1} & \textbf{1} & \textbf{1} & 0.94 & 0.94 & \textbf{1} & 1.13 \\ \cline{3-11} 
 &  & 200 & 0.76 & 0.95 & 0.95 & 0.95 & \textbf{1} & \textbf{1} & \textbf{1} & 1.05 \\ \cline{2-11} 
 & \multirow{3}{*}{35} & 5 & \textbf{1} & 0.82 & 0.82 & 0.82 & 0.82 & 0.82 & \textbf{1} & 1.45 \\ \cline{3-11} 
 &  & 50 & 0.87 & 0.67 & \textbf{1} & 0.93 & 0.80 & 0.80 & \textbf{1} & 1.20 \\ \cline{3-11} 
 &  & 200 & 0.82 & \textbf{1} & \textbf{1} & \textbf{1} & \textbf{1} & \textbf{1} & \textbf{1} & 1.24 \\ \hline
\end{tabular}
\caption{Success rates normalized relative to the SPO.}
\label{tab:policies-success-rate-ratios}
\end{table*}

\begin{figure*}[t]
    \centering  
    \begin{subfigure}[b]{0.325\textwidth}
        \centering
        \includegraphics[width=\textwidth]{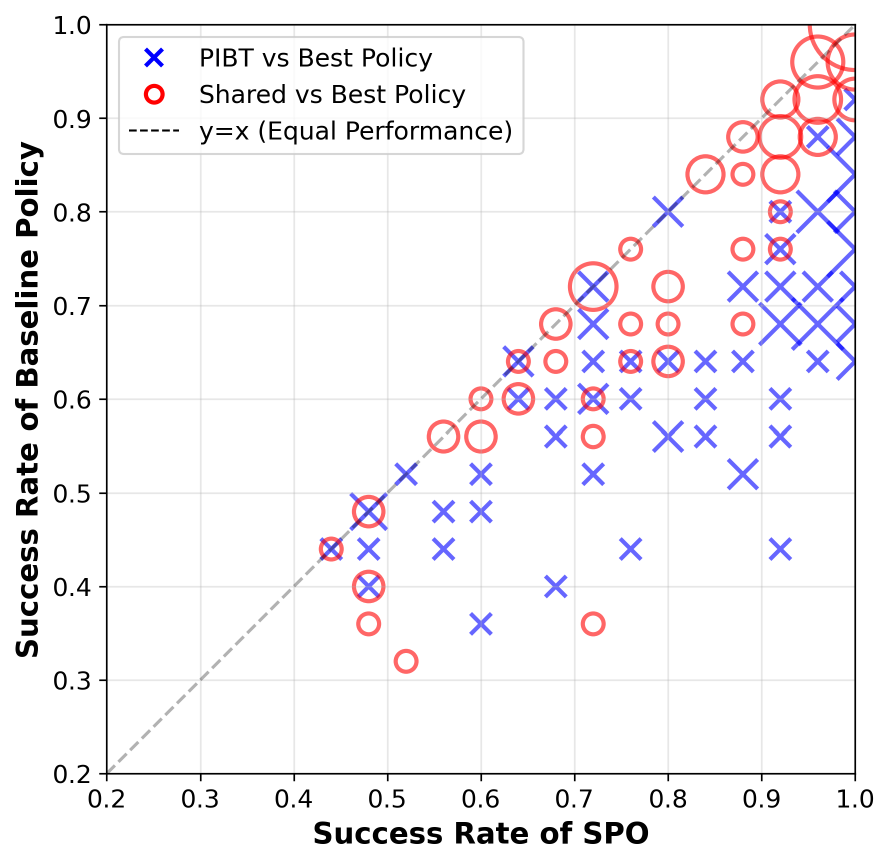}
        \caption{Success Rate Comparison for Room}
        \label{fig:room-success-rate}
    \end{subfigure}   
    \begin{subfigure}[b]{0.325\textwidth}
        \centering
        \includegraphics[width=\textwidth]{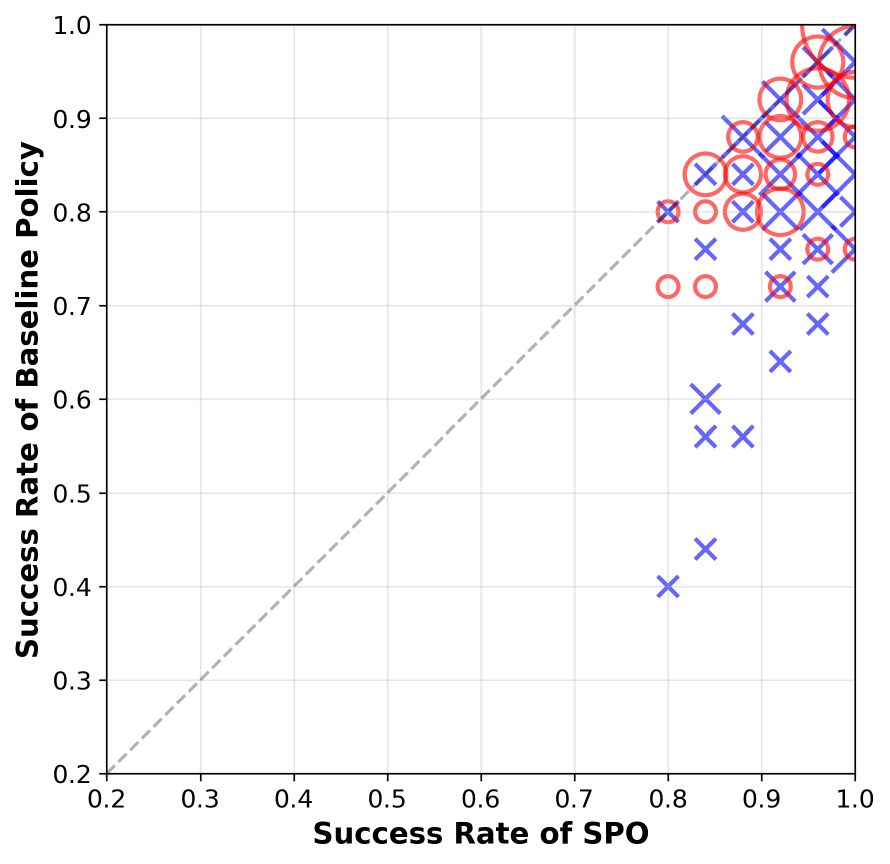}
        \caption{Success Rate Comparison for Random}
        \label{fig:random-success-rate}
    \end{subfigure}
    \begin{subfigure}[b]{0.325\textwidth}
        \centering
        \includegraphics[width=\textwidth]{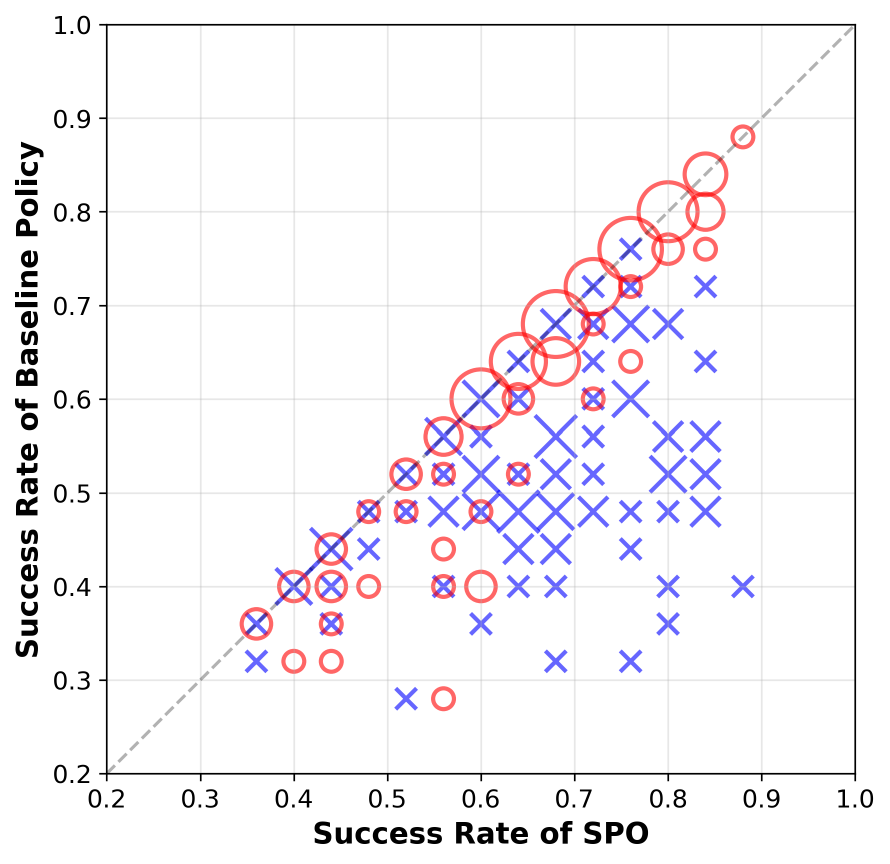}
        \caption{Success Rate Comparison for Maze}
        \label{fig:maze-success-rate}
    \end{subfigure} \quad
    \caption{Comparison of success rate of the baselines against the SPO}
    \label{fig:comparison-success-rate}
\end{figure*}

For evaluating the overall success rates of the baselines against the SPO algorithm, in Fig.~\ref{fig:comparison-success-rate}, each data point depicts the comparative success rate of one of the baselines versus the SPO for a single combination of execution window and budget.

In Fig.~\ref{fig:comparison-success-rate} the line y=x represents equal results for the baselines and the SPO, every data point below the line represents advantage for the SPO over the baseline. To visualize overlapping data, the marker size indicates the multiplicity of results at a given coordinate; larger markers signify that multiple parameter configurations achieved the same success rate. As expected, no data point in above the equal line, this is because the SPO can also choose the baselines when they have the advantage over the budgeted policies.

In the Room map results we notice an advantage towards the shared baseline than the PIBT. We observe that although there are some data points on the equal line, most of the data points are below it showing a clear advantage for the SPO over the baselines. Furthermore, the Shared policy appears to outperform PIBT. The maximum performance discrepancy against SPO for the Shared policy occurred at the data point [0.72, 0.36], whereas the largest divergence for PIBT was observed at [0.92, 0.44].
In the Random map, we observe generally higher success rates; however, PIBT exhibits a larger performance deficit relative to the SPO compared to the Shared policy. Despite these elevated success rates, the SPO maintains a distinct advantage over the baselines. Specifically, the maximum performance discrepancy for the Shared policy occurred at [1.00, 0.76], whereas the largest divergence for PIBT was significantly wider, observed at [0.80, 0.40].
In the Maze map we see both the baselines and the SPO struggle to get high success rates in contrast to the other maps. The Shared policy retains an advantage over PIBT in this map environment. The maximum performance discrepancy for the Shared policy was observed at [0.56, 0.28], whereas PIBT exhibited a notably larger divergence at [0.88, 0.40].

\subsection{Results 2: Solution Quality}
\begin{figure*}[t]
    \centering
    \begin{subfigure}[b]{0.325\textwidth}
        \centering
        \includegraphics[width=\textwidth]{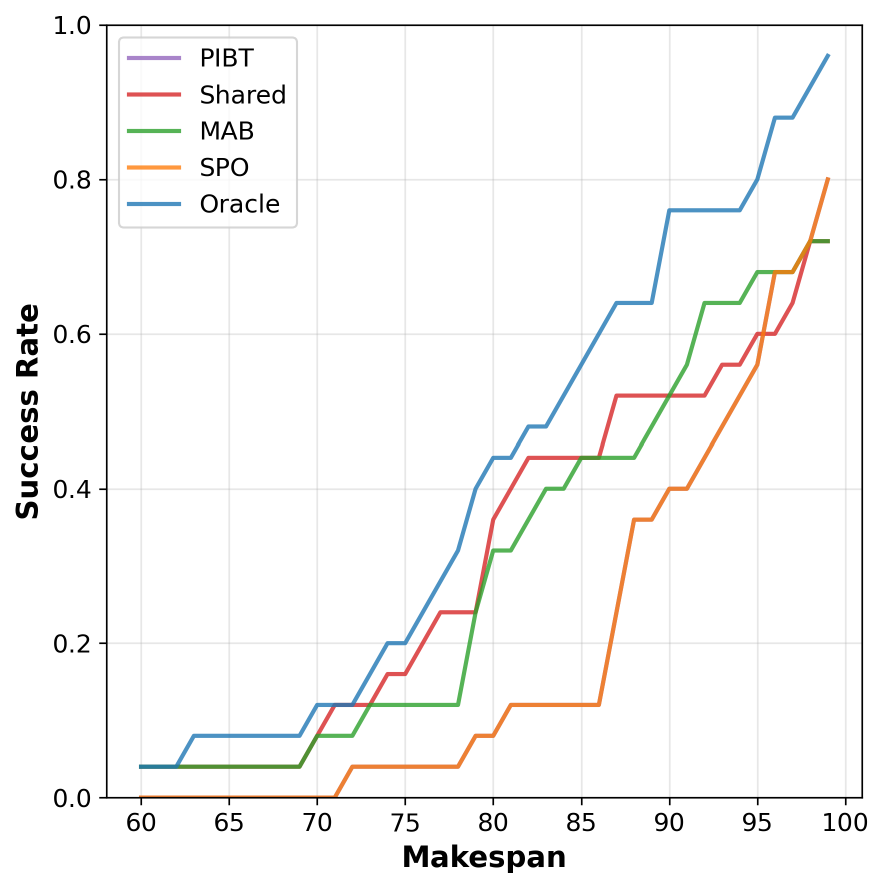}
        \caption{Cactus plot for the min AUC ratio for Room map at prefix=5 and budget=15}
        \label{fig:min-cactus-plot}
        \label{fig:max-cactus-plot}
    \end{subfigure}
    \begin{subfigure}[b]{0.325\textwidth}
        \centering
        \includegraphics[width=\textwidth]{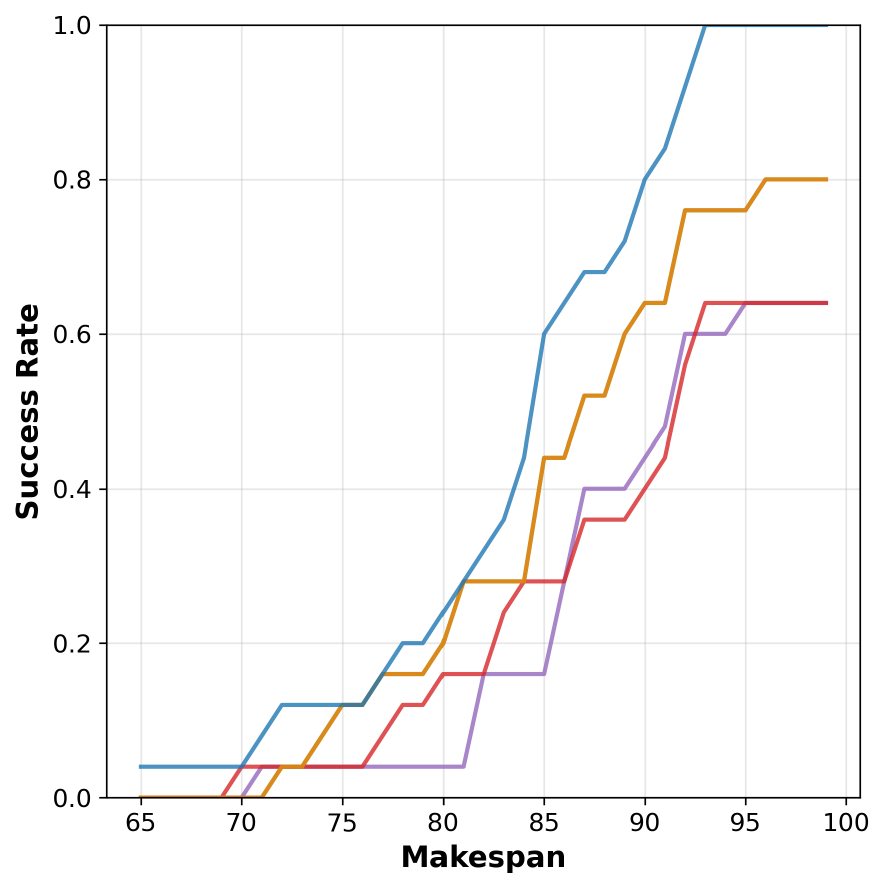}
        \caption{Cactus plot for the max AUC ratio for Room map at prefix=10 and budget=15}
        \label{fig:max-cactus-plot}
    \end{subfigure}
    \caption{Example of Cactus plots of the MAX and MIN AUC}
    \label{fig:cactus-plot-example}
\end{figure*}

Since the final success rate provides limited insight on the algorithm performance, we will also consider the quality of the solutions. This is done by looking at the area under the curve (AUC) of a cactus plot, which an example of can be seen in Fig.~\ref{fig:cactus-plot-example}. In this plot, the x-axis represents the allowable makespan (ranging from the minimum makespan found to 99), while the y-axis denotes the normalized success rate (0 to 1). Each data point indicates the percentage of scenarios that would be considered successful if the makespan upper bound were set to that specific value. The AUC metric is obtained by summing the value of each data point and this indicates the algorithm's overall performance profile, capturing both its ability to find high-quality solutions and its reliability in solving harder instances. In the example both the configuration with the maximum AUC and with the minimum AUC for the Room map are shown. It should be noted that in Fig.~\ref{fig:min-cactus-plot}, the SPO plot completely overlaps the PIBT baseline, as the SPO identifies PIBT as the best performing policy. Likewise, in Fig.~\ref{fig:max-cactus-plot}, the SPO line lies directly on top of the MAB policy line for the same reason.

In Table \ref{tab:auc-ratio} we compare the ability of the SPO to find higher quality results against the baselines. Each cell displays the ratio between the AUC of the SPO and the AUC of the best-performing baseline for a given execution window and budget. Thus, values greater than 1 indicate that the SPO outperformed the baselines, while values less than 1 imply superior performance by the baselines. Unlike the overall success rate, where the SPO is optimal by definition, it can be outperformed in the AUC metric because the SPO is designed to select the policy with the highest success rate, not the highest AUC.

\begin{table*}[]
\centering
\begin{tabular}{|c|ccccccccccccccc|}
\hline
Map & \multicolumn{1}{c|}{Prefix} & \multicolumn{1}{c|}{5} & \multicolumn{1}{c|}{10} & \multicolumn{1}{c|}{15} & \multicolumn{1}{c|}{20} & \multicolumn{1}{c|}{25} & \multicolumn{1}{c|}{50} & \multicolumn{1}{c|}{80} & \multicolumn{1}{c|}{100} & \multicolumn{1}{c|}{150} & \multicolumn{1}{c|}{200} & \multicolumn{1}{c|}{250} & \multicolumn{1}{c|}{300} & \multicolumn{1}{c|}{350} & 400 \\ \hline
\multirow{7}{*}{Room} & \multicolumn{1}{c|}{5} & \multicolumn{1}{c|}{0.89} & \multicolumn{1}{c|}{1} & \multicolumn{1}{c|}{\textbf{1.09}} & \multicolumn{1}{c|}{1} & \multicolumn{1}{c|}{\textbf{1.21}} & \multicolumn{1}{c|}{1} & \multicolumn{1}{c|}{\textbf{1.01}} & \multicolumn{1}{c|}{\textbf{1.12}} & \multicolumn{1}{c|}{\textbf{1.13}} & \multicolumn{1}{c|}{\textbf{1.06}} & \multicolumn{1}{c|}{1} & \multicolumn{1}{c|}{0.99} & \multicolumn{1}{c|}{0.96} & \textbf{1.05} \\ \cline{2-16} 
 & \multicolumn{1}{c|}{10} & \multicolumn{1}{c|}{1} & \multicolumn{1}{c|}{\textbf{1.05}} & \multicolumn{1}{c|}{\textbf{1.16}} & \multicolumn{1}{c|}{\textbf{1.09}} & \multicolumn{1}{c|}{1} & \multicolumn{1}{c|}{1} & \multicolumn{1}{c|}{\textbf{1.02}} & \multicolumn{1}{c|}{\textbf{1.01}} & \multicolumn{1}{c|}{1} & \multicolumn{1}{c|}{0.98} & \multicolumn{1}{c|}{\textbf{1.01}} & \multicolumn{1}{c|}{1} & \multicolumn{1}{c|}{0.94} & \textbf{1.15} \\ \cline{2-16} 
 & \multicolumn{1}{c|}{15} & \multicolumn{1}{c|}{0.84} & \multicolumn{1}{c|}{1} & \multicolumn{1}{c|}{1} & \multicolumn{1}{c|}{1} & \multicolumn{1}{c|}{1} & \multicolumn{1}{c|}{\textbf{1.02}} & \multicolumn{1}{c|}{\textbf{1.06}} & \multicolumn{1}{c|}{\textbf{1.05}} & \multicolumn{1}{c|}{1} & \multicolumn{1}{c|}{1} & \multicolumn{1}{c|}{\textbf{1.07}} & \multicolumn{1}{c|}{\textbf{1.01}} & \multicolumn{1}{c|}{\textbf{1.02}} & \textbf{1.06} \\ \cline{2-16} 
 & \multicolumn{1}{c|}{20} & \multicolumn{1}{c|}{1} & \multicolumn{1}{c|}{1} & \multicolumn{1}{c|}{1} & \multicolumn{1}{c|}{\textbf{1.09}} & \multicolumn{1}{c|}{\textbf{1.10}} & \multicolumn{1}{c|}{0.91} & \multicolumn{1}{c|}{\textbf{0.98}} & \multicolumn{1}{c|}{\textbf{1.10}} & \multicolumn{1}{c|}{\textbf{1.03}} & \multicolumn{1}{c|}{\textbf{1.01}} & \multicolumn{1}{c|}{1} & \multicolumn{1}{c|}{\textbf{1.04}} & \multicolumn{1}{c|}{1} & 1 \\ \cline{2-16} 
 & \multicolumn{1}{c|}{25} & \multicolumn{1}{c|}{\textbf{1.06}} & \multicolumn{1}{c|}{1} & \multicolumn{1}{c|}{\textbf{1.11}} & \multicolumn{1}{c|}{\textbf{1.20}} & \multicolumn{1}{c|}{\textbf{1.17}} & \multicolumn{1}{c|}{\textbf{1.16}} & \multicolumn{1}{c|}{\textbf{1.12}} & \multicolumn{1}{c|}{\textbf{1.13}} & \multicolumn{1}{c|}{0.98} & \multicolumn{1}{c|}{\textbf{1.06}} & \multicolumn{1}{c|}{\textbf{1.10}} & \multicolumn{1}{c|}{\textbf{1.06}} & \multicolumn{1}{c|}{\textbf{1.02}} & \textbf{1.10} \\ \cline{2-16} 
 & \multicolumn{1}{c|}{30} & \multicolumn{1}{c|}{0.97} & \multicolumn{1}{c|}{\textbf{1.08}} & \multicolumn{1}{c|}{\textbf{1.05}} & \multicolumn{1}{c|}{1.12} & \multicolumn{1}{c|}{\textbf{1.19}} & \multicolumn{1}{c|}{\textbf{1.13}} & \multicolumn{1}{c|}{\textbf{0.96}} & \multicolumn{1}{c|}{\textbf{1.02}} & \multicolumn{1}{c|}{\textbf{1.16}} & \multicolumn{1}{c|}{\textbf{1.11}} & \multicolumn{1}{c|}{1} & \multicolumn{1}{c|}{1} & \multicolumn{1}{c|}{1} & 1 \\ \cline{2-16} 
 & \multicolumn{1}{c|}{35} & \multicolumn{1}{c|}{1} & \multicolumn{1}{c|}{\textbf{1.03}} & \multicolumn{1}{c|}{\textbf{1.11}} & \multicolumn{1}{c|}{1} & \multicolumn{1}{c|}{1} & \multicolumn{1}{c|}{\textbf{1.16}} & \multicolumn{1}{c|}{1} & \multicolumn{1}{c|}{\textbf{1.07}} & \multicolumn{1}{c|}{0.96} & \multicolumn{1}{c|}{\textbf{1.09}} & \multicolumn{1}{c|}{1} & \multicolumn{1}{c|}{1} & \multicolumn{1}{c|}{1} & 1 \\ \hline
 &  &  &  &  & \textbf{} &  & \textbf{} &  & \textbf{} &  &  &  &  &  &  \\ \hline
\multirow{7}{*}{Random} & \multicolumn{1}{c|}{5} & \multicolumn{1}{c|}{1} & \multicolumn{1}{c|}{\textbf{1.03}} & \multicolumn{1}{c|}{1} & \multicolumn{1}{c|}{\textbf{1.14}} & \multicolumn{1}{c|}{\textbf{1.24}} & \multicolumn{1}{c|}{\textbf{1.06}} & \multicolumn{1}{c|}{1} & \multicolumn{1}{c|}{0.98} & \multicolumn{1}{c|}{1} & \multicolumn{1}{c|}{\textbf{1.01}} & \multicolumn{1}{c|}{\textbf{1.06}} & \multicolumn{1}{c|}{\textbf{1.01}} & \multicolumn{1}{c|}{1} & \textbf{1.20} \\ \cline{2-16} 
 & \multicolumn{1}{c|}{10} & \multicolumn{1}{c|}{0.99} & \multicolumn{1}{c|}{1} & \multicolumn{1}{c|}{\textbf{1.03}} & \multicolumn{1}{c|}{\textbf{1.07}} & \multicolumn{1}{c|}{\textbf{1.09}} & \multicolumn{1}{c|}{\textbf{1.03}} & \multicolumn{1}{c|}{\textbf{1.09}} & \multicolumn{1}{c|}{\textbf{1.06}} & \multicolumn{1}{c|}{1} & \multicolumn{1}{c|}{1} & \multicolumn{1}{c|}{\textbf{1.01}} & \multicolumn{1}{c|}{\textbf{1.01}} & \multicolumn{1}{c|}{\textbf{1.03}} & 1 \\ \cline{2-16} 
 & \multicolumn{1}{c|}{15} & \multicolumn{1}{c|}{1} & \multicolumn{1}{c|}{0.94} & \multicolumn{1}{c|}{\textbf{1.01}} & \multicolumn{1}{c|}{\textbf{1.08}} & \multicolumn{1}{c|}{\textbf{1.02}} & \multicolumn{1}{c|}{\textbf{1.03}} & \multicolumn{1}{c|}{\textbf{1.08}} & \multicolumn{1}{c|}{\textbf{1.03}} & \multicolumn{1}{c|}{\textbf{1.05}} & \multicolumn{1}{c|}{1} & \multicolumn{1}{c|}{1.06} & \multicolumn{1}{c|}{\textbf{1.05}} & \multicolumn{1}{c|}{1} & \textbf{1.04} \\ \cline{2-16} 
 & \multicolumn{1}{c|}{20} & \multicolumn{1}{c|}{1} & \multicolumn{1}{c|}{\textbf{1.27}} & \multicolumn{1}{c|}{1} & \multicolumn{1}{c|}{1} & \multicolumn{1}{c|}{\textbf{1.10}} & \multicolumn{1}{c|}{1} & \multicolumn{1}{c|}{\textbf{1.02}} & \multicolumn{1}{c|}{\textbf{1.07}} & \multicolumn{1}{c|}{0.90} & \multicolumn{1}{c|}{1} & \multicolumn{1}{c|}{\textbf{1.04}} & \multicolumn{1}{c|}{1} & \multicolumn{1}{c|}{1} & 0.97 \\ \cline{2-16} 
 & \multicolumn{1}{c|}{25} & \multicolumn{1}{c|}{\textbf{1.02}} & \multicolumn{1}{c|}{\textbf{1.05}} & \multicolumn{1}{c|}{1} & \multicolumn{1}{c|}{1} & \multicolumn{1}{c|}{\textbf{1.17}} & \multicolumn{1}{c|}{\textbf{1.08}} & \multicolumn{1}{c|}{1} & \multicolumn{1}{c|}{\textbf{1.05}} & \multicolumn{1}{c|}{0.92} & \multicolumn{1}{c|}{1} & \multicolumn{1}{c|}{1} & \multicolumn{1}{c|}{0.78} & \multicolumn{1}{c|}{1} & 1 \\ \cline{2-16} 
 & \multicolumn{1}{c|}{30} & \multicolumn{1}{c|}{\textbf{1.01}} & \multicolumn{1}{c|}{\textbf{1.47}} & \multicolumn{1}{c|}{\textbf{1.06}} & \multicolumn{1}{c|}{1} & \multicolumn{1}{c|}{\textbf{1.15}} & \multicolumn{1}{c|}{1} & \multicolumn{1}{c|}{\textbf{1.07}} & \multicolumn{1}{c|}{\textbf{1.11}} & \multicolumn{1}{c|}{\textbf{1.03}} & \multicolumn{1}{c|}{0.86} & \multicolumn{1}{c|}{0.91} & \multicolumn{1}{c|}{1} & \multicolumn{1}{c|}{1} & \textbf{1.11} \\ \cline{2-16} 
 & \multicolumn{1}{c|}{35} & \multicolumn{1}{c|}{\textbf{1.09}} & \multicolumn{1}{c|}{\textbf{1.11}} & \multicolumn{1}{c|}{\textbf{1.05}} & \multicolumn{1}{c|}{1} & \multicolumn{1}{c|}{\textbf{1.09}} & \multicolumn{1}{c|}{\textbf{1.13}} & \multicolumn{1}{c|}{\textbf{1.05}} & \multicolumn{1}{c|}{\textbf{1.10}} & \multicolumn{1}{c|}{\textbf{1.04}} & \multicolumn{1}{c|}{\textbf{1.08}} & \multicolumn{1}{c|}{0.96} & \multicolumn{1}{c|}{0.92} & \multicolumn{1}{c|}{1} & \textbf{1.11} \\ \hline
 &  &  &  &  &  &  &  &  &  &  &  &  &  &  &  \\ \hline
\multirow{7}{*}{Maze} & \multicolumn{1}{c|}{5} & \multicolumn{1}{c|}{1} & \multicolumn{1}{c|}{\textbf{1.02}} & \multicolumn{1}{c|}{1} & \multicolumn{1}{c|}{\textbf{1.08}} & \multicolumn{1}{c|}{\textbf{1.02}} & \multicolumn{1}{c|}{1} & \multicolumn{1}{c|}{\textbf{1.09}} & \multicolumn{1}{c|}{1} & \multicolumn{1}{c|}{1} & \multicolumn{1}{c|}{1} & \multicolumn{1}{c|}{1} & \multicolumn{1}{c|}{1} & \multicolumn{1}{c|}{1} & \textbf{1.01} \\ \cline{2-16} 
 & \multicolumn{1}{c|}{10} & \multicolumn{1}{c|}{1} & \multicolumn{1}{c|}{\textbf{1.05}} & \multicolumn{1}{c|}{\textbf{1.02}} & \multicolumn{1}{c|}{\textbf{1.01}} & \multicolumn{1}{c|}{1} & \multicolumn{1}{c|}{\textbf{1.02}} & \multicolumn{1}{c|}{1} & \multicolumn{1}{c|}{1} & \multicolumn{1}{c|}{1} & \multicolumn{1}{c|}{1} & \multicolumn{1}{c|}{1} & \multicolumn{1}{c|}{1} & \multicolumn{1}{c|}{\textbf{1.05}} & \textbf{1.06} \\ \cline{2-16} 
 & \multicolumn{1}{c|}{15} & \multicolumn{1}{c|}{1} & \multicolumn{1}{c|}{1} & \multicolumn{1}{c|}{1} & \multicolumn{1}{c|}{1} & \multicolumn{1}{c|}{\textbf{1.08}} & \multicolumn{1}{c|}{\textbf{1.01}} & \multicolumn{1}{c|}{1} & \multicolumn{1}{c|}{1} & \multicolumn{1}{c|}{\textbf{1.04}} & \multicolumn{1}{c|}{1} & \multicolumn{1}{c|}{\textbf{1.06}} & \multicolumn{1}{c|}{1} & \multicolumn{1}{c|}{1} & 1 \\ \cline{2-16} 
 & \multicolumn{1}{c|}{20} & \multicolumn{1}{c|}{1} & \multicolumn{1}{c|}{1} & \multicolumn{1}{c|}{1} & \multicolumn{1}{c|}{1} & \multicolumn{1}{c|}{\textbf{1.12}} & \multicolumn{1}{c|}{\textbf{1.03}} & \multicolumn{1}{c|}{\textbf{1.04}} & \multicolumn{1}{c|}{1} & \multicolumn{1}{c|}{1} & \multicolumn{1}{c|}{\textbf{1.02}} & \multicolumn{1}{c|}{\textbf{1.01}} & \multicolumn{1}{c|}{1} & \multicolumn{1}{c|}{1} & \textbf{1.02} \\ \cline{2-16} 
 & \multicolumn{1}{c|}{25} & \multicolumn{1}{c|}{1} & \multicolumn{1}{c|}{1} & \multicolumn{1}{c|}{0.72} & \multicolumn{1}{c|}{1} & \multicolumn{1}{c|}{0.91} & \multicolumn{1}{c|}{\textbf{1.05}} & \multicolumn{1}{c|}{1} & \multicolumn{1}{c|}{1} & \multicolumn{1}{c|}{1} & \multicolumn{1}{c|}{1} & \multicolumn{1}{c|}{1} & \multicolumn{1}{c|}{1} & \multicolumn{1}{c|}{1} & 1 \\ \cline{2-16} 
 & \multicolumn{1}{c|}{30} & \multicolumn{1}{c|}{1} & \multicolumn{1}{c|}{1} & \multicolumn{1}{c|}{1} & \multicolumn{1}{c|}{1} & \multicolumn{1}{c|}{1} & \multicolumn{1}{c|}{\textbf{1.04}} & \multicolumn{1}{c|}{\textbf{1.07}} & \multicolumn{1}{c|}{\textbf{1.01}} & \multicolumn{1}{c|}{1} & \multicolumn{1}{c|}{1} & \multicolumn{1}{c|}{1} & \multicolumn{1}{c|}{1} & \multicolumn{1}{c|}{1} & 1 \\ \cline{2-16} 
 & \multicolumn{1}{c|}{35} & \multicolumn{1}{c|}{1} & \multicolumn{1}{c|}{1} & \multicolumn{1}{c|}{0.85} & \multicolumn{1}{c|}{\textbf{1.06}} & \multicolumn{1}{c|}{1} & \multicolumn{1}{c|}{\textbf{1.28}} & \multicolumn{1}{c|}{1} & \multicolumn{1}{c|}{\textbf{1.06}} & \multicolumn{1}{c|}{\textbf{1.25}} & \multicolumn{1}{c|}{\textbf{1.01}} & \multicolumn{1}{c|}{0.97} & \multicolumn{1}{c|}{1} & \multicolumn{1}{c|}{1} & \textbf{1.03} \\ \hline
\end{tabular}
\caption{Ratios between the AUC of the SPO and the AUC of the best-performing baseline for the all maps}
\label{tab:auc-ratio}
\end{table*}

In the Table ~\ref{tab:auc-ratio} we observe the overall results of the SPO against the best performing baseline across all the configurations across all maps. For the  Room map, we see a clear advantage in using the SPO compared to a baseline: 48 permutations show an advantage equal or greater than 1\% for the SPO, while only 31 permutations show an advantage equal or greater than 1\% for the baselines. We can also notice that the baselines had better results when the budget was very low compared to the execution window. The highest SPO advantage in this table sits at 34\% when the execution window and budget per agent are 10 and 15 respectively, while the highest baseline advantage sits at 38\% when the execution window and budget per agent are 5 and 15 respectively.

We also observe the overall results of the SPO against the best performing baseline across all the configurations in the Random map. We see a clear advantage in using the SPO compared to the baseline: 55 permutations show an advantage equal or greater than 1\% for the SPO, while only 11 permutations show an advantage equal or greater than 1\% for the baselines. The highest SPO advantage in this table sits at 47\% when the execution window and budget per agent are 30 and 11 respectively, while the highest baseline advantage sits at 22\% when the execution window and budget per agent are 25 and 300 respectively.

Lastly, we observe the overall results of the SPO against the best performing baseline across all the configurations in the Maze map. We still see an advantage in using the SPO compared to a baseline: 29 permutations show an advantage equal or greater than 1\% for the policies, while only 4 permutations show an advantage greater than equal or 1\% for the baselines. We can also notice that the baselines had equal or better results in a triangle shape that represent an area where the budget is low compared to the execution window. We also observe an area with roughly the same results with the SPO and the baselines (ratio is 1) when there is a much larger budget compared to the execution window. The highest SPO advantage in this table sits at 28\% when the execution window and budget per agent are 35 and 50 respectively, while the highest baseline advantage is similar at 26\% when the execution window and budget per agent are 35 and 5 respectively.

In the Room and Maze maps, the SPO typically yields results similar to or worse than the baselines in scenarios with large execution windows relative to the budget. However, this trend is reversed in the Random map, where the SPO achieves its most significant advantage under these exact parameters. This deviation stems from two factors: the use of SIPPS and the nature of the map's difficulty, which arises from agent density rather than static obstacle density. Unlike standard A*, which expands a separate node for each time step during a wait, SIPPS represents waiting as a single interval, significantly reducing the search space in high-density scenarios. This efficiency conserves the budget, allowing the allocation policies to utilize the remaining resources strategically. Consequently, this mechanism shifts instances with high execution windows and low budgets into a more manageable difficulty tier. High agent density also negatively affects PIBT. While capable of escaping difficult situations, PIBT generates paths of poor quality. Consequently, when the budget is too tight to allow for solution improvement, the overall performance of the PIBT baseline suffers significantly.

\subsection{Results 3: SPO Ablation}
In addition, to determine which policies exert the most significant influence on the SPO, we conducted an ablation analysis.
The SPO utilizes six distinct policies: CPB, RCPB, PID, MAB, Shared, and PIBT. To quantify the marginal contribution of each policy, we employed an iterative backward elimination strategy. In each iteration, we evaluated the performance impact of excluding each remaining policy individually. The policy whose exclusion resulted in the least significant performance degradation was permanently removed from the set. This process was repeated recursively until a single policy remained, allowing us to rank the policies by their impact on the SPO's overall success.

\begin{table}[]
\centering
\begin{tabular}{|c|c|c|c|c|c|}
\hline
\multirow{2}{*}{Room} & RCPB & Shared & PIBT & MAB & PID \\ \cline{2-6} 
 & 1 (4\%) & 3 (12\%) & 7 (20\%) & 21 (20\%) & 47 (28\%) \\ \hline
\multirow{2}{*}{Random} & RCPB & PIBT & CPB & Shared & PID \\ \cline{2-6} 
 & 4 (4\%) & 6 (8\%) & 12 (4\%) & 25 (12\%) & 45 (16\%) \\ \hline
\multirow{2}{*}{Maze} & Shared & RCPB & MAB & CPB & PIBT \\ \cline{2-6} 
 & 0 (0\%) & 1 (4\%) & 3 (4\%) & 18 (12\%) & 30 (28\%) \\ \hline
\end{tabular}
\caption{Number of parameter configurations affected by cumulative policy exclusion relative to the baseline SPO.}
\label{tab:spo-ablation}
\end{table}

Table~\ref{tab:spo-ablation} illustrates the divergence between the baseline SPO and the cumulatively ablated versions. In the Room map, for instance, RCPB demonstrated the least impact, with its exclusion altering the result of only one parameter configuration. The subsequent removal of Shared ,resulting in the exclusion of RCPB + Shared, increased the discrepancy to three configurations. This trend continues until only CPB remains; as the final surviving policy with the most significant impact on SPO performance, it is not listed among the excluded sets.
These results corroborate our earlier observation that no single policy is universally superior. For instance, the importance of PIBT is environment-dependent: in the Random map, it is the second policy to be eliminated (small impact), whereas in the Maze map, it ranks as the second most critical policy. Despite these variations, a general trend emerges: the RCPB policy is consistently excluded in the early stages, while PID is reliably retained until the final iterations, indicating its high contribution to the SPO.

\subsection{Overall Results}
The overall results demonstrate a distinct advantage for the SPO over the baselines, evidenced by superior performance in both success rates and AUC metrics. Furthermore, our analysis confirms that no single policy is universally dominant, as efficacy varies with problem parameters. This variability underscores the necessity of the SPO framework, which dynamically adapts to maximize algorithmic performance.
Across all map environments, the baselines demonstrate superior performance in scenarios characterized by either extreme resource scarcity or abundance. When the budget is insufficient relative to the execution window, it proves more efficient to generate a rapid initial solution via PIBT and dedicate the minimal remaining resources to \lnsone. Conversely, when the budget is plentiful, the overhead of intelligent allocation yields diminishing returns, rendering the proposed policies unnecessary. The SPO mainly outperforms the baselines in the middle range of budget availability, situated between the lower and upper extremes. In these scenarios, simply applying a static policy is inefficient; instead, an intelligent allocation of the budget is required to maximize performance. By dynamically adjusting to these constraints, the SPO is able to achieve higher success rates than the standard baseline approaches.

\section{Conclusion and Future Work}
In this paper, we studied the Real-Time MAPF problem (RT-MAPF), which is a MAPF problem where every planning period must finish within a fixed, small, time budget, after which the agents must commit to performing a predefined sequence of moves (the execution window). We show that in RT-MAPF it is crucial to intelligently allocate the planning budget as opposed to simply running existing MAPF algorithms and halting them when the planning budget is exhausted. 
We proposed several ways to distribute the planning budget between the agents for the state-of-the-art MAPF algorithm \lnstwo. 
In particular, we propose \confprop \revconfprop and \PID, which are methods for computing how much budget a neighborhood of agents should be given within \lnstwo. Furthermore, we introduced a real-time Multi-Armed Bandit mechanism for dynamic policy adaptation. Our analysis confirms that optimal performance varies by scenario; since no single policy is best, the implementation of an Oracle-like selection algorithm is essential. Our experimental results show that when using a smart budget allocation policy we are able to move the agents towards their targets faster.
Future work may consider incorporating online learning mechanisms to adjust the budget allocated to different agents during execution, and applying the different budget allocation policies in lifelong MAPF settings. 


\bibliographystyle{IEEEtran}
\bibliography{main.bib}

\end{document}